\documentclass[10pt, aps, pra]{revtex4-2}
\usepackage{orcidlink}
\usepackage[T1]{fontenc}
\usepackage[utf8]{inputenc}
\usepackage{babel}
\usepackage{braket}
\usepackage{xcolor}
\usepackage{natbib}
\usepackage{amsfonts, amsmath, amssymb}
\usepackage{graphicx}
\usepackage[top = 2cm, left = 2.5cm, right = 2.5cm, bottom = 2cm]{geometry}
\usepackage{hyperref}
\hypersetup{
    colorlinks = true,
    allcolors = blue,
    pdftitle = {Dynamics and Spectral Response of linear-quadratic optomechanical interaction: Effects of pure dephasing},
}

\newcommand{\mi}{\mathrm{i}}
\newcommand{\re}{\mathrm{Re}}
\newcommand{\im}{\mathrm{Im}}

\begin{document}
\title{Dynamics and Spectral Response of linear-quadratic optomechanical interaction: Effects of pure dephasing}
\author{Alejandro R. Urz\'ua\,\orcidlink{0000-0002-6255-5453}}
\email{alejandro@icf.unam.mx}
\affiliation{Instituto de Ciencias F\'isicas, Universidad Nacional Aut\'onoma de M\'exico\\ Av. Universidad sn, Col. Chamilpa, 62210, Cuernavaca, Morelos, M\'exico}

\date{\today}

\begin{abstract}
    In this manuscript, the decoherence dynamics and spectral response of an optomechanical system, with linear and quadratic couplings, is addressed. The decoherence considered arises from pure dephasing, described by the Milburn stochastic evolution of the Schr\"odinger equation. In the first part of the manuscript, it is shown how the decoherence rate influences the evolution of the number of phonons, and the quadrature of the mechanical resonator. In the second part of the manuscript, an attempt to look at the spectral response of the mechanical part of the system is given using non-stationary spectroscopy. It is emphasized the response of the resonator in its equilibrium position when a single photon excitation in the cavity field is prepared. Coherent states are also considered in the cavity field and the mechanical resonator. Results and discussion comparing the inclusions of the linear, quadratic, and linear-quadratic couplings are given.
\end{abstract}

\maketitle

\section{Introduction}

Quantum optomechanics has been an active field of research for at least the last three decades. It is not coincidental, or a casualty, since there's great interest in obtaining a profound understanding of how systems of different orders of magnitude relates when they are coupled through some designed interaction; leading to use it as technological platforms to test quantum effects, and measurements in the nanoscale \cite{Barzanjeh2022OptomechanicsTechnologies}. In the most broad sense, an optomechanical system is composed of a photonic (light) and phononic (matter) subsystems, interacting through a proper degree of freedom. It is of interest when the field, due to radiation pressure exerts a displacement in the boundary of a cavity, amplified and feedbacked if the boundary can be displaced in some manner, harmonically in the most of the cases. Nowadays, there exists a very comprehensive literature on the fundamentals of the optomechanical realm \cite{2014CavityOptomechanics,Aspelmeyer2014CavityOptomechanics,Kippenberg2008CavityMesoscale, Milburn2011AnOptomechanics}. In the traditional approximation, the frequency of the cavity field is modified by the displacement of the mechanical boundary due to this radiation pressure, thus creating a feedback loop between how much the displacement is, and how much conversion between quanta of light and matter takes place due dynamical Casimir effect \cite{Ferreri2022InterplayEffect,DelGrosso2019PhotonSystem,Law1994EffectiveMedium}. At the heart of this phenomenon, when cavity QED is used to describe it, lays an approximation to the displacement of the boundary. This approximation states that for a resonant cavity of length $L$, when the instantaneous displacement $x$ is orders of magnitude tinier, we can expand a quotient in a series of contributions, typically keeping to the first order when the number of quanta in the field coupled to the linear displacement of the cavity boundary, leading to a completely integrable system. Because the expansion of the interaction term, in incremental orders, it naturally motivates to ask whether their contributions modify the behavior of the system. Despite the first order of interaction is highly nonlinear, it can be linearized using the polaron transformation, a technique not suitable for higher orders due to their disadvantage of disregarding non-vanishing terms, like the quadratic and upper. Archetypical examples of linear and quadratic couplings are the well-known Fabry-P\'erot cavity, and the “membrane-in-the-middle” cavity \cite{Thompson2008StrongMembrane,Law1995InteractionFormulation,Gigan2006Self-coolingPressure}, respectively. Although the linear coupling can be diminished with a well-engineered cavity, the quadratic term is persistent, thus the interest to studying the expansion of the position cavity frequency up-to second order interaction. There exists a plethora of results involving linear and quadratic interaction in optomechanical systems: the so-called optical spring \cite{Rai2008QuantumSpring}, single-photon mechanical control \cite{Nunnenkamp2011Single-photonOptomechanics}, photon blockade in linear and quadratic couplings \cite{Rabl2011PhotonSystems}, mechanical cooling \cite{Gu2015MechanicalNonlinearity, Liao2013PhotonSystems}, implementation of quadratic couplings in circuits \cite{Kim2015CircuitOptomechanics}, preparation of nonclassical states \cite{Brunelli2018UnconditionalOptomechanics}, squeezing-enhanced quantum sensing \cite{Zhang2024Squeezing-enhancedOptomechanics}, quantum signatures \cite{Machado2019QuantumOptomechanics}, mechanical induced transparency \cite{Zhang2018OptomechanicallyCoupling}, to say something about it. Having knowledge of these results, and stating that the losing rates of the cavity and the resonator are slower compared with other quantities in the system, we can also put in the picture the decoherence effects due dephasing in the evolution of the system. Since the Hamiltonian carries the energetic information on the system, the time-evolution scheme will give us a chance to model how it evolves under specific conditions. In the traditional approximation, where dispersive-dissipative cavity is considered, quantum Langevin equation are settled to deal with the losses and driving of the system \cite{Barchielli2015QuantumSystems, Khorasani2018MethodOptomechanics, Marquardt2008QuantumCooling, Singh2014QuantumOscillator}. Taking an alternative approach, the stochastic evolution of the Schr\"odinger equation proposed by Milburn \cite{Milburn1991IntrinsicMechanics}, can handle in some manner related to quantum master equation, an evolution damped by a decoherence control parameter, which will preserve the total number of excitations, as it is devised in a non-dissipative manner. This process, called \emph{intrinsic decoherence}, has been the subject of numerous studies since its introduction, due to the ability to handle damping in the systems. Despite the former discussion about the backgrounds of the proposal \cite{Finkelstein1993CommentMechanics, Milburn1993Reply}, and recently inquiries about the suitability of this kind of decoherence as a physical theory \cite{Danelli2024AreTheories}, we can justify the election as an alternative to modelling a master equation with the Hamiltonian itself as decaying operator \cite{Kuang1997Jaynes-CummingsDamping}. As a direct antecedent, the investigations in coupled and displaced harmonic oscillator, along simple moving-mirror system have been done \cite{RUrzua2022IntrinsicOscillator, Urzua2024IntrinsicInteraction, Urzua2024MovingDecoherence}, but a large corpus of results exists whiting the literature \cite{Moya-Cessa1993IntrinsicInteraction, Mousavi2024DifferentFormalism, Love2008IntrinsicCondensate, Mohanty1997IntrinsicSystems, Wu2017IntrinsicSystems, Essakhi2022IntrinsicDynamics, Muthuganesan2021IntrinsicNonlocality, AitChlih2021DynamicsInteraction,Mohamed2020GenerationEffects, Abdel-Aty2021NonlocalityDecoherence, Hab-arrih2023VirtualDecoherence}. As a final ingredient in this manuscript, it is considered the use of the well-developed non-stationary spectroscopy proposed Eberly-W\'odkiewicz \cite{Eberly1977TheLight}, which has been successfully used for many kinds of system and processes; pure atomic \cite{Roman-Ancheyta2018Time-dependentShelving, Castro-Beltran2025Time-dependentBeats, delosSantos-Sanchez2022Strain-spectroscopyQubits}, atom-field \cite{Villanueva-Vergara2020EffectModel, Medina-Dozal2024SpectralModel}, and thermalization process \cite{Roman-Ancheyta2019SpectralThermalization,Salado-Mejia2021SpectroscopyRegime}. The advantage of capturing information about the energetic transitions and resonances in the system as a function of time is notable, and whose limit in time is the well-known Wiener–Khinchin power spectrum, it leads then to a natural choosing to explore the energetic setting of the system evolution in time. Conjuring the system, the pure dephasing evolution scheme and the spectroscopic visualization process, the proposal in this manuscript it's to ask how the dynamics of an optomechanical system, with linear and quadratic couplings, and experiencing a damping due to phase decoherence is experienced, and how is the spectral response in the evolution of the mechanical system. Particularly, it is described the single-photon influence in the resonator, plus a comparison when the cavity field and the resonator are prepared initially in coherent states as well.

The manuscript is organized as follows: In Sec.~\ref{sec:dynamics} we start revisiting the fundamentals of the quantum optomechanical Hamiltonian formulation up to a second order expansion in the interaction, keeping in mind that the treatment is platform-independent as far as we don't specify the interdependency of the coupling strengths. It follows the solution given by diagonalization using unitary transformations. Expectation values measurements are given to discuss the dynamics of the evolution, and how the phase damping induces cooling in the mechanical resonator. In Sec.~\ref{sec:spectral} we use the non-stationary spectral resolving method of Eberly-W\'odkiewicz to determine how the inclusion of the second order interaction modifies the spectral response, and to gain insights about the evolution of energetic transitions when the decoherence ceases the quanta conversion between the cavity field and the mechanical resonator. Finally, a concluding section \ref{sec:conclusions} is given to discuss and highlight the findings.

\section{Dynamics}\label{sec:dynamics}
\begin{figure}[htbp]
    \centering
    \includegraphics[width = 0.5\linewidth]{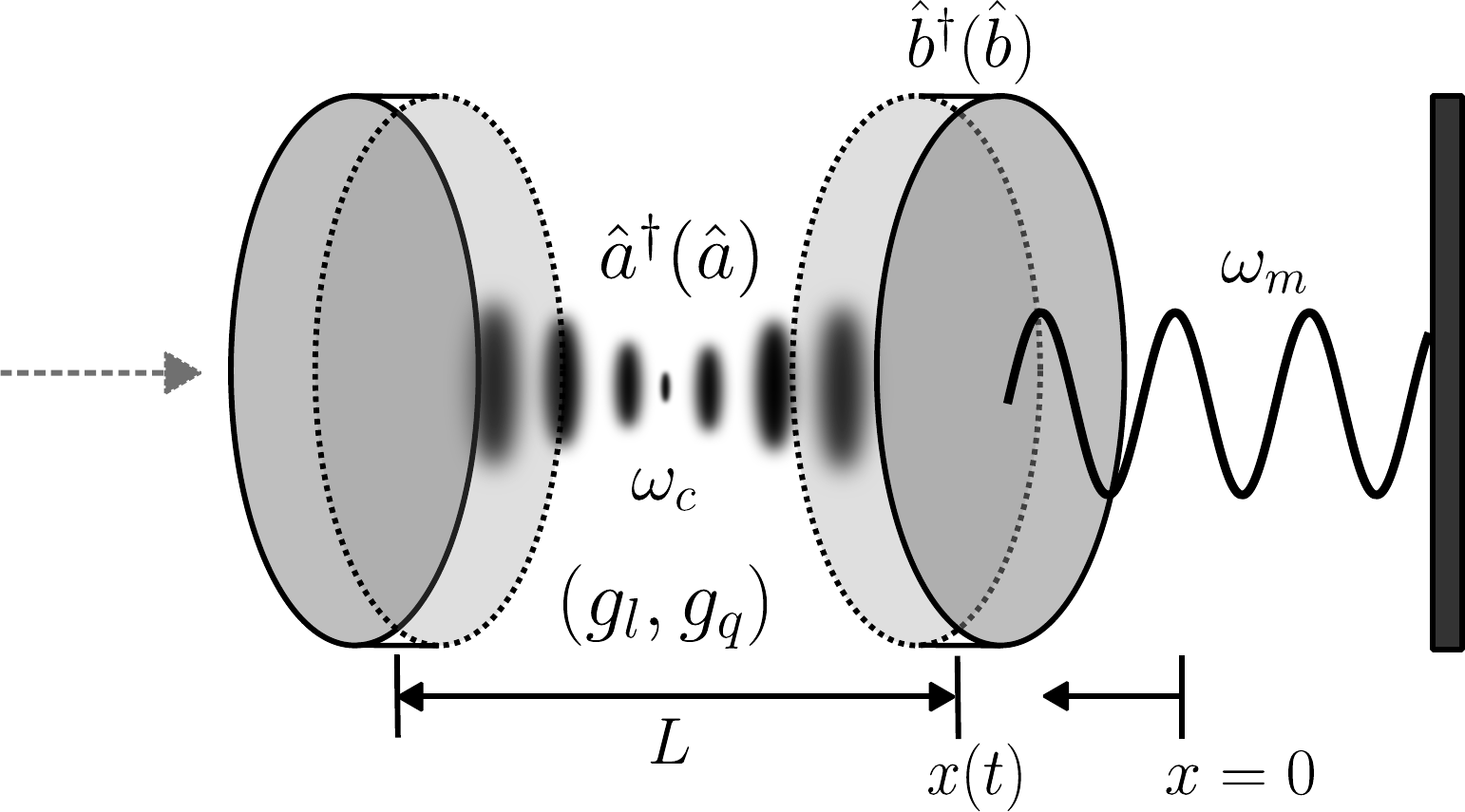}
    \caption{Schematic representation of an optomechanical cavity sustaining an electromagnetic field $\hat{a}^{\dagger}(\hat{a})$ with frequency $\omega_{c}$, which is coupled to a mechanical resonator (reflecting moving mirror) $\hat{b}^{\dagger}(\hat{b})$ with frequency $\omega_{m}$. The coupling is denoted as a function of the linear and quadratic contributions $g_{l}$ and $g_{q}$.}
    \label{fig:FP_cav}
\end{figure}
The system under study can be thought at first as the quintessential optomechanical cavity, where an electromagnetic field is enclosed by two reflective boundaries, one of then, the mechanical resonator, experienced length displacement $L(x) = L - x$ due to radiation pressure \cite{Kippenberg2007CavityOpto-Mechanics, Vahala2003OpticalMicrocavities}, where $L$ is the initial length of the cavity, and $x\propto\hat{b}^{\dagger} + \hat{b}$ the instantaneous position of the moving boundary, as the scheme depicted in Fig.~\ref{fig:FP_cav}. In the standard treatment, this length determines the optical frequencies $\omega(x) = 2\pi c/L(x)$ coupled to the cavity field $\hat{n}$. Suppose the mechanical resonator has a natural frequency $\omega_{m}$, the interacting system at large is given by $\omega(x)\hat{n} + \omega_{m}\hat{N}$. Here the usual commutation relations from bosonic algebra for the field and the mirror take place, $[\hat{a}, \hat{a}^{\dagger}] = [\hat{b}, \hat{b}^{\dagger}$] = 1. If we expand the denominator of the optical frequency in terms of $\hat{x}\ll L$, where $\hat{x}$ represents the quantized version of the mirror position, we arrive at a series of contributions
\begin{equation}
    \omega(\hat{x}) = \omega_{c}\sum\limits_{k = 0}^{\infty}\left(\frac{L}{\hat{x}}\right)^{k},
\end{equation}
where $\omega_{c}$ is the proper frequency of the cavity field. This leads us to consider, along the fundamental term, the two-relevant $k = 1~\textrm{and}~2$, representing linear and quadratic contributions to the optical frequency modification due to radiation feedback on-and-from the mechanical resonator. Finally, let's define the Hamiltonian of the system under these considerations as
\begin{equation}\label{eq:H}
    \hat{H} = \hat{H}_{\texttt{free}} + \hat{H}_{\texttt{int}}^{(l)} + \hat{H}_{\texttt{int}}^{(q)}
\end{equation}
where \texttt{free} and \texttt{int}, are the free and linear-quadratic interaction Hamiltonians, explicitly given by
\begin{equation}\label{eq:H_f-i}
\begin{gathered}
    \hat{H}_{\texttt{free}} = \omega_{c}\hat{n} + \omega_{m}\hat{N}\\
    \hat{H}_{\texttt{int}}^{(l)} = - g_{l} \hat{n}\left(\hat{b}^{\dagger} + \hat{b}\right),\quad \hat{H}_{\texttt{int}}^{(q)} =   g_{q} \hat{n}\left(\hat{b}^{\dagger} + \hat{b}\right)^{2},
\end{gathered}
\end{equation}
in which the interaction strength $g_{l}$ and $g_{q}$ will be platform-independent, allowing different realizations of the Hamiltonian, like optomechanical cavities or superconducting circuits \cite{Kippenberg2007CavityOpto-Mechanics, Kim2015CircuitOptomechanics}.

\paragraph*{Diagonalization.} This system has been addressed previously, using Lie algebraic methods to determine its temporal evolution when linear strong coupling is considered \cite{MedinaDozal2023}, also the temporal evolution of the linear-quadratic coupling when a forcing term in the cavity field is introduced \cite{MedinaDozal2024}. In this work, the diagonalization of equation \eqref{eq:H}-\eqref{eq:H_f-i} is done using unitary transformations, a one-mode squeezing and a one-mode displacement, in the mechanical resonator basis $\{\hat{b}, \hat{b}^{\dagger}\}$. Thus, applying the following operators
\begin{equation}
    S(r(\hat{n})) = \exp\left[\frac{1}{2}r(\hat{n})\left(\hat{b}^{\dagger}{}^{2} - \hat{b}^{2}\right)\right],\quad D(\alpha(\hat{n})) = \exp\left[\alpha(\hat{n})\left(\hat{b}^{\dagger} - \hat{b}\right)\right],
\end{equation}
we arrive at the diagonalized (\texttt{diag}) Hamiltonian,
\begin{equation}\label{eq:Hdiag}
\begin{aligned}
	\hat{H}_{\texttt{diag}} &= \omega_{c}\hat{n} + \bar{\omega}_{m}(\hat{n})\hat{N} - \bar{\omega}_{m}(\hat{n})\vert\alpha(\hat{n})\vert^{2}\\
    &= \omega_{c}\hat{n} + \bar{\omega}_{m}(\hat{n})\hat{N} - \frac{g_{l}^{2}\hat{n}^{2}e^{-2r(\hat{n})}}{\bar{\omega}_{m}(\hat{n})},
\end{aligned}
\end{equation}
where they emerge quantities that depend linearly on the number of photons $\hat{n}$, carrying the information about the field inside the cavity, these are
\begin{equation}
    r(\hat{n}) = \frac{1}{4}\ln\left\vert 1 + \frac{4g_{q}\hat{n}}{\omega_{m}}\right\vert,\quad \bar{\omega}_{m}(\hat{n}) = \sqrt{\omega_{m}\left(\omega_{m} + 4g_{q}\hat{n}\right)},\quad \alpha(\hat{n}) = -\frac{g_{l}\hat{n}e^{-r}}{\bar{\omega}_{m}(\hat{n})}.
\end{equation}
Note as well the appearance of a quadratic number of photons operator $\hat{n}^{2}$, this can be interpreted as a Kerr-like medium inside the cavity, which is responsible for non-linear effects.

Having the diagonalized expression \eqref{eq:Hdiag}, we can return to the original reference frame using the inverse transformations,
\begin{equation}\label{eq:Hinv}
\begin{aligned}
    \hat{H}_{\texttt{diag}} &\equiv D(\alpha(\hat{n}))D(r(\hat{n}))\,\hat{H}\,S^{\dagger}(r(\hat{n}))D^{\dagger}(\alpha(\hat{n}))\\
    &\Rightarrow \hat{H} = S^{\dagger}(r(\hat{n}))D(\alpha(\hat{n}))\,\hat{H}_{\texttt{diag}}\,D(\alpha(\hat{n}))S(r(\hat{n})).
\end{aligned}
\end{equation}

A complete survey on the diagonalization procedure can be found in appendix \ref{app:diag}. In Figure~\ref{fig:eigen_diag}, it is shown the eigenvalues of \eqref{eq:Hdiag}, defined as $\mathcal{E}_{n,N} = \omega_{c}n + \bar{\omega}_{m}(n)N - \bar{\omega}_{m}(n)\vert\alpha(n)\vert^{2}$, for $n, N\in\mathbb{N}$. Panels A and B show the energy levels when $\omega_{c} = \omega_{m}$, fixing $g_{q}$ and variating $g_{l}$ and vice versa, respectively. We can see that the only uncoupled non-degenerate level is $\mathcal{E}_{0,0}$; this degeneracy of levels is broken as $g_{l}$ grows in panel A, but it is broken as soon as $g_{q} \neq 0$ in panel B. On the other hand, panels C and D show the energy levels when $\omega_{m} \ll \omega_{c}$, fixing one of the couplings and variating the other. Here the degeneracy of levels is totally broken from the start, showing crossings at some specific values of $g_{l}$ and $g_{q}$. Note that fixing $g_{q}$ and variating $g_{l}$ in panel C can lead to a triple level crossing.\\
\begin{figure}
    \centering
    \includegraphics[width = \linewidth]{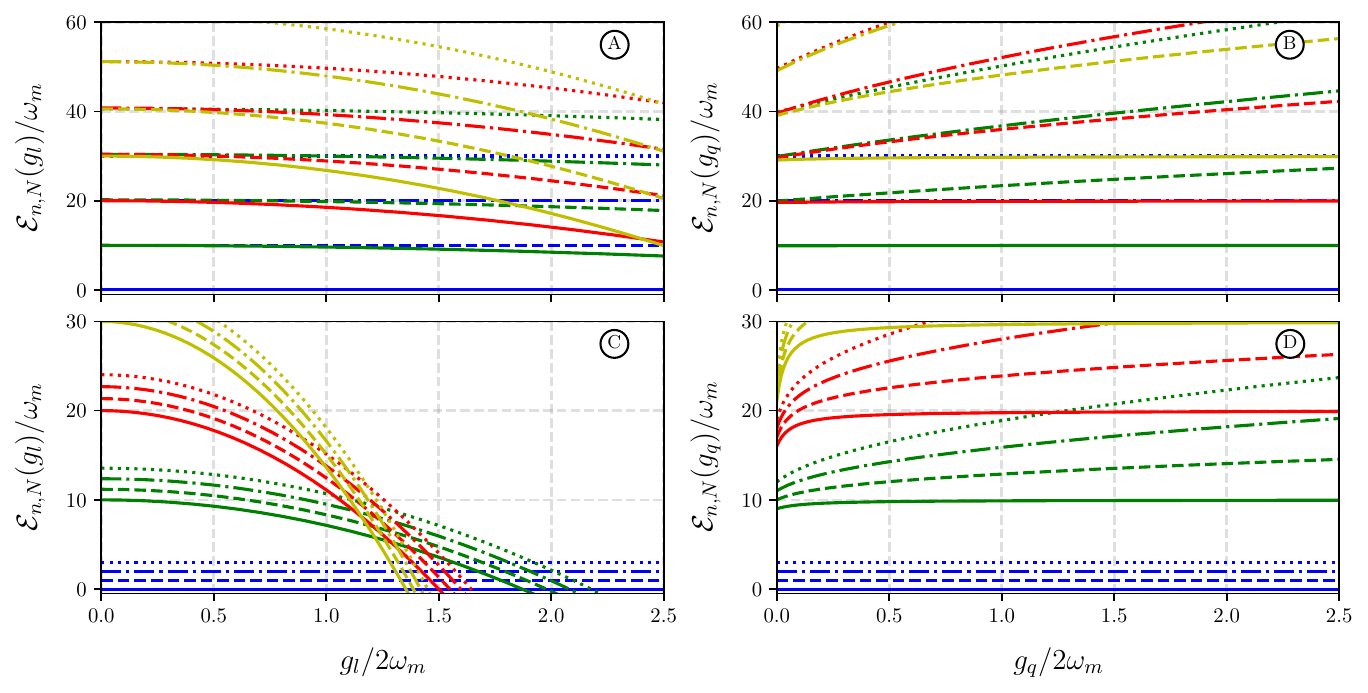}
    \caption{
    Eigenvalues $\mathcal{E}_{n, N}(g_{l}, g_{q})$ of the Hamiltonian \eqref{eq:H_f-i} as function of the pair photon-phonon $(n, N)$, and optomechanical couplings, $g_{l}$ and $g_{q}$. Blue, green, red, and yellow correspond for $n = \{0, 1, 2, 3\}$, respectively; whereas, solid, dashed, dashed-dot, and dotted correspond for $N = \{0, 1, 2, 3\}$, respectively. It is plotting against, either $g_{l}$ in left column, or $g_{q}$ in right column. Panel A. $\omega_{m} = \omega_{c}$, $g_{q} = 0.01$. Panel B. $\omega_{m} = \omega_{c}$, $g_{l} = 0.1$. Panel C. $\omega_{m} = 0.1\omega_{c}$, $g_{q} = 0.01$. Panel D. $\omega_{m} = 0.1\omega_{c}$, $g_{l} = 0.1$. 
    }
    \label{fig:eigen_diag}
\end{figure}

\paragraph*{Decoherence.} Now, according to Milburn \cite{Milburn1991IntrinsicMechanics}, we can define the evolution of a wavefunction in the intrinsic decoherence framework by
\begin{equation}\label{eq:rho_t}
    \hat{\rho}_{\gamma}(t) = e^{-\gamma t}\sum\limits_{k = 0}^{\infty}\frac{\left(\gamma t\right)^{k}}{k!} \left(\ket{\psi_{k}}\bra{\psi_{k}}\right)_{\gamma},
\end{equation}
where the $k$-th ket is given by
\begin{equation}\label{eq:psik}
    \ket{\psi_{k}}_{\gamma} = e^{-\frac{\mi k}{\gamma}\hat{H}}\ket{\psi(0)},
\end{equation}
in which the quantity $\gamma$ represents the rate of decoherence, determining the slope of the damping in the excitation-conserving evolution of the systems' phase. We can also write down the evolution operator of the decoherence as
\begin{equation}\label{eq:U_op}
    \hat{\mathcal{U}}_{\gamma}(t) = e^{-\gamma t}\sum\limits_{k = 0}^{\infty}\frac{(\gamma t)^{k}}{k!}~\hat{\mathcal{V}}\left(\tfrac{k}{\gamma}\right),
\end{equation}
where $\hat{\mathcal{V}}(k/\gamma)$ is the $k$-th component of the evolution operator, given by
\begin{equation}\label{eq:V_op}
    \hat{\mathcal{V}}\left(\tfrac{k}{\gamma}\right) = \hat{S}^{\dagger}(r(\hat{n}))\hat{D}^{\dagger}(\alpha(\hat{n}))~e^{-\frac{\mi k}{\gamma}\hat{H}_{\texttt{diag}}}~\hat{S}(r(\hat{n}))\hat{D}(\alpha(\hat{n})).
\end{equation}
A rise of warning, must be placed here. Since an infinite sum must be done to obtain the final expression in time, we need to perform first the respective action on the intended operator, and then the sum. Previously, an attempt to describe the evolution of the linear coupling optomechanical interaction \cite{Urzua2024MovingDecoherence} settles a precedent on what we can expect, at least looking at the mirror's response to the phase damping.

Since the Hamiltonian \eqref{eq:H_f-i} determines the energetic disposition of the system, the first interest is to determine the time evolution of the dynamical quantities, like the number of phonons and position quadrature of the mirror, parametrically dependent on the decoherence rate $\gamma$. This election is justified by the fact that the evolution operator of the decoherence $\eqref{eq:U_op}$ via the $\hat{H}_{\texttt{diag}}$, commutes with the number of photons operator $\hat{n}$, so looking at the response of the mirror to the cavity field, seems appropriate at this point. Let's start calculating the mechanical bosonic operators in the $(k/\gamma)$-representation given by the application of \eqref{eq:V_op}, to this matter we need to determine the representation of the bilateral action
\begin{equation}
    \hat{\mathcal{O}}\left(\tfrac{k}{\gamma}\right) = \hat{\mathcal{V}}^{\dagger}\left(\tfrac{k}{\gamma}\right)~\hat{\mathcal{O}}~\hat{\mathcal{V}}\left(\tfrac{k}{\gamma}\right).
\end{equation}
The calculation involves five steps where we reverse the representation to the original frame of reference. Acting over $\{\hat{b}^{\dagger}, \hat{b}\}$, we obtain
\begin{equation}
\begin{aligned}
    \hat{b}^{\dagger}\left(\tfrac{k}{\gamma}\right) &= \hat{b}^{\dagger}\,\chi_{1}^{(n)}\left(\tfrac{k}{\gamma}\right) + \hat{b}\,\chi_{2}^{(n)}\left(\tfrac{k}{\gamma}\right) + \chi_{3}^{(n)}\left(\tfrac{k}{\gamma}\right)\\
    \hat{b}\left(\tfrac{k}{\gamma}\right) &= \hat{b}\,\chi_{1}^{(n)}\left(\tfrac{k}{\gamma}\right)^{*} + \hat{b}^{\dagger}\,\chi_{2}^{(n)}\left(\tfrac{k}{\gamma}\right)^{*} + \chi_{3}^{(n)}\left(\tfrac{k}{\gamma}\right)^{*},
\end{aligned}
\end{equation}
where the companion coefficients are
\begin{equation}\label{eq:chi_coeffs}
\begin{aligned}
    \chi_{1}^{(n)}\left(\tfrac{k}{\gamma}\right) &= \cosh^{2}~r(\hat{n})~e^{\frac{\mi k\bar{\omega}_{m}(\hat{n})}{\gamma}} - \sinh^{2}~r(\hat{n})~e^{-\frac{\mi k\bar{\omega}_{m}(\hat{n})}{\gamma}}\\
    \chi_{2}^{(n)}\left(\tfrac{k}{\gamma}\right) &= \frac{1}{2}\left(e^{\frac{\mi k\bar{\omega}_{m}(\hat{n})}{\gamma}} - e^{-\frac{\mi k\bar{\omega}_{m}(\hat{n})}{\gamma}}\right)\sinh~2r(\hat{n})\\
    \chi_{3}^{(n)}\left(\tfrac{k}{\gamma}\right) &= \alpha(\hat{n})\left(\cosh~r(\hat{n})~e^{\frac{\mi k\bar{\omega}_{m}(\hat{n})}{\gamma}} - \sinh~r(\hat{n})~e^{\frac{-\mi k\bar{\omega}_{m}(\hat{n})}{\gamma}} - e^{-r(\hat{n})}\right).
\end{aligned}
\end{equation}
When each of these coefficients are summed up according to \eqref{eq:U_op}, and the limit $\gamma\rightarrow\infty$ is taken, the expressions reduce to those obtained under the usual Heisenberg evolution. Compare it with those in the appendix \eqref{eq:exp_coeffs}. Having these expressions, we can compute the time-evolved operator for the number of phonons, 
\begin{equation}\label{eq:Nmean}
\begin{aligned}
    \hat{N}_{\gamma}(t) &= 
    \hat{N}\left(\zeta_{1}^{(n)}(t) + \zeta_{2}^{(n)}(t)\right) + \hat{b}^{\dagger}{}^{2}\,\zeta_{12}^{(n)}(t) + \hat{b}^{2}\,\zeta_{12}^{(n)}(t)^{*}\\
    & + \hat{b}^{\dagger}\left(\zeta_{13}^{(n)}(t) + \zeta_{23}^{(n)}(t)^{*}\right) + \hat{b}\left(\zeta_{23}^{(n)}(t) + \zeta_{13}^{(n)}(t)\right) + \zeta_{2}^{(n)}(t) + \zeta_{3}^{(n)}(t),
\end{aligned}
\end{equation}
where the companion terms are given by
\begin{equation}
    \zeta_{l}^{(n)}(t) = e^{-\gamma t}\sum\limits_{k = 0}^{\infty}\frac{(\gamma t)^{k}}{k!}~\Big\vert\chi_{l}^{(n)}\left(\tfrac{k}{\gamma}\right)\Big\vert^{2},\quad \zeta_{lm}^{(n)}(t) = e^{-\gamma t}\sum\limits_{k = 0}^{\infty}\frac{(\gamma t)^{k}}{k!}~\chi_{l}^{(n)}\left(\tfrac{k}{\gamma}\right)\chi_{m}^{(n)}\left(\tfrac{k}{\gamma}\right)^{*}.
\end{equation}
And the position quadrature
\begin{equation}\label{eq:Xmean}
    \hat{X}_{\gamma}(t) = \hat{b}^{\dagger}\left(\nu_{1}^{(n)}(t) + \nu_{2}^{(n)}(t)^{*}\right) + \hat{b}\left(\nu_{1}^{(n)}(t)^{*} + \nu_{2}^{(n)}(t)\right) + 2\re\left\{\nu_{3}^{(n)}(t)\right\}, 
\end{equation}
where the companion terms are given by
\begin{equation}\label{eq:nu_coeffs}
    \nu_{l}^{(n)}(t) = e^{-\gamma t}\sum\limits_{k = 0}^{\infty}\frac{(\gamma t)^{k}}{k!}~\chi_{l}^{(n)}\left(\tfrac{k}{\gamma}\right).
\end{equation}

\begin{figure}[t]
    \centering
    \includegraphics[width = \linewidth]{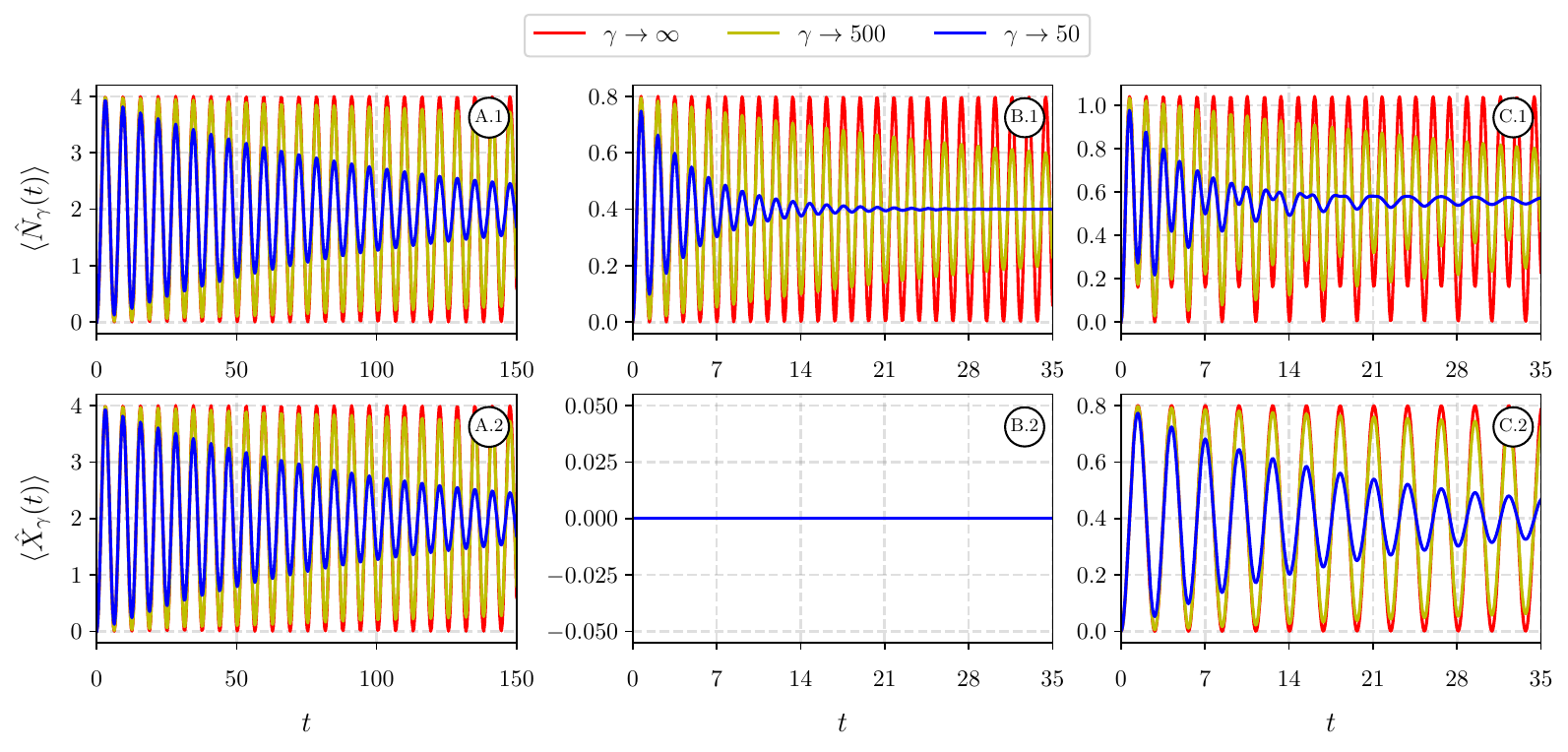}
    \caption{$\hat{N}_{\gamma}(t)$ and $\hat{X}_{\gamma}(t)$. $\{A,B,C\}.\square$ are evolutions with linear, quadratic and linear-quadratic couplings, respectively. Setting $\omega_{c} = \omega_{m} = 1.0$, $g_{l} = g_{q} = 1.0$ and $\ket{\psi(0)}_{\mathrm{num}} = \ket{1, 0}$. We see how the decoherence rate $\gamma$ influences in the diminishing of the amplitude in the three regimes.}
    \label{fig:NXnum_expect}
\end{figure}

\begin{figure}[t]
    \centering
    \includegraphics[width = \linewidth]{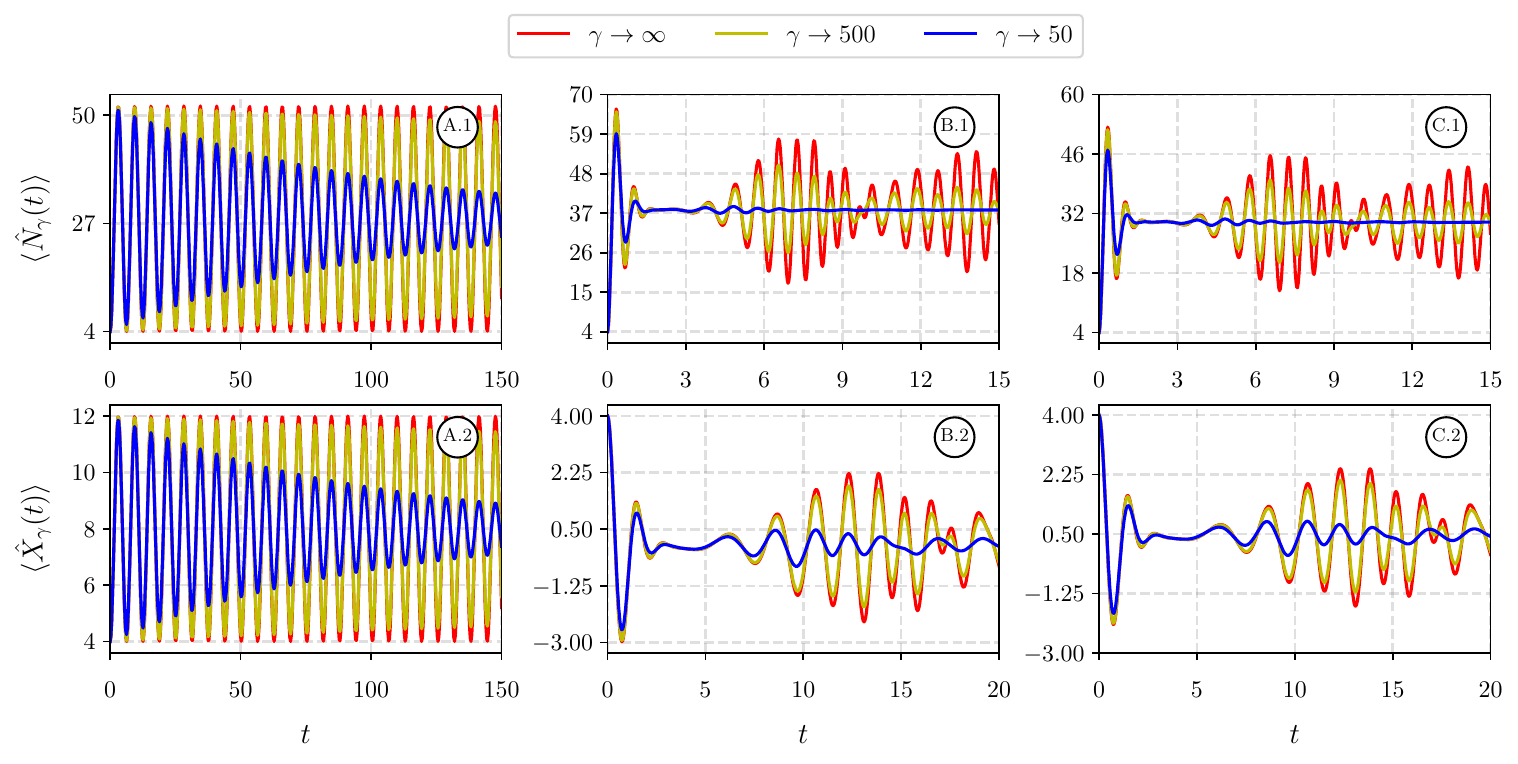}
    \caption{$\hat{N}_{\gamma}(t)$ and $\hat{X}_{\gamma}(t)$. $\{A,B,C\}.\square$ are evolutions with linear, quadratic and linear-quadratic couplings, respectively. Setting $\omega_{c} = \omega_{m} = 1.0$, $g_{l} = g_{q} = 1.0$ and $\ket{\psi(0)}_{\mathrm{coh}} = \ket{2, 2}$. We see how the decoherence rate $\gamma$ influences in the diminishing of the amplitude in the three regimes.}
    \label{fig:NXcoh_expect}
\end{figure}

In the previous set of equations, the superscript $(n)$ denotes a direct play of the number of photons in the cavity field, modifying each term where it appears. Moreover, the frequency of the field is not explicitly apparent in none of the dynamical expressions for the mechanical resonator. Now, it is just a matter of selecting a suitable initial condition. For this, we can start looking at the behavior of single excitation in the cavity field \cite{Nunnenkamp2011Single-photonOptomechanics, Gu2015MechanicalNonlinearity}, searching for insights about the transfer of radiation pressure to the mirror in the presence of linear, quadratic and linear-quadratic coupling, parametrically dependent on the decoherence rate $\gamma$. Setting number states in the field and the mirror, $\ket{n, m}$, with just a single photon $\bar{n} = 1$ and no phonons $\bar{N} = 0$, this is $\ket{\psi(0)}_{\mathrm{num}} = \ket{1}_{c}\otimes\ket{0}_{m}$. In Fig.~\ref{fig:NXnum_expect} we see how the system evolves under different circumstances: $A.1$ and $A.2$ is a linear only coupling, $g_{q} = 0$, where the number of phonons and the position quadrature are the same, thus the one photon radiation pressure is back and forth between the mirror and the cavity; turning on the decoherence rate, induces a monotonically decreasing of the amplitude, leading to a mean phonon number and quadrature position of $\bar{N} = 2$ and $\bar{X} = 2$, respectively. $B.1$  and $B.2$ is a quadratic only coupling, $g_{l} = 0$, here the number of phonons behaves in almost the same way as the linear coupling, but accelerating the decreasing of amplitude converging to a mean value of $\bar{N} = 0.4$, on the other hand, the position quadrature is always null, $\bar{X}(t) = 0\;\forall t$, meaning that the mirror isn't experiencing back and fort due radiation pressure, the initial photon is elastically bouncing to a fixed boundary. Finally, $C.1$ and $C.2$ accounts for the presence of both, the linear and quadratic couplings simultaneously, where we see in the number of phonons the presence of double modulation, indicating squeezing of states, on the other hand, the position quadrature behaves very clean, modulating a decreasing amplitude converging to a mean value.
To end this section, we also look at the evolution considering coherent states, in the cavity field and the mirror. For this, we can set the initial condition using the coherent representation of states in the mirror, and the photon distribution in the field, $\ket{\alpha, \beta}_{\mathrm{coh}} = e^{-\alpha^{2}}\sum_{k = 0}^{\infty}(\alpha^{2k}/k!)~\ket{n, \beta}$. Taking this, and using \eqref{eq:Nmean} and \eqref{eq:Xmean}, we can obtain the respective behavior, which is shown in Fig.~\ref{fig:NXcoh_expect}, observing the same label system. Here, a well-known feature of the presence of a quadratic coupling is the revival-like of mean quantities, showed by the optical spring of Rai and Agarwal \cite{Rai2008QuantumSpring}. We see that the decoherence has an effect of cooling the mirror, thus leading to a mean phonon value $\bar{N}$, without chance of more revival like cycles. 

\section{Spectral response}\label{sec:spectral}
Obtaining the spectrum of the mechanical resonator is straight-forward, since the two-time correlation enables us to use the coefficients given by the terms $\nu_{l}^{(n)}(t)$ in equation \eqref{eq:nu_coeffs}, which are self-dependent, meaning that $(\chi_{l}^{(n)}(k/\gamma))^{*}\rightarrow(\nu_{l}^{(n)}(t))^{*}$. The time-dependent spectrum is defined by \cite{Eberly1977TheLight}, 
\begin{equation}
    \mathcal{S}_{\gamma}(\Gamma, \omega; t) = 2\Gamma e^{-2\Gamma t}\int_{0}^{t}~dt_{1}~e^{(\Gamma - \mi\omega)t_{1}}~\int_{0}^{t}~dt_{2}~e^{(\Gamma + \mi\omega)t_{2}}~\braket{\hat{b}_{\gamma}^{\dagger}(t_{1})\hat{b}_{\gamma}(t_{2})},
\end{equation}
where the correlation function for an arbitrary initial condition $\ket{\psi(0)}$ is
\begin{equation}
\begin{aligned}
    \braket{\hat{b}_{\gamma}^{\dagger}(t_{1})\hat{b}_{\gamma}(t_{2})} = \bra{\psi(0)}\Big[&\hat{N}\left[\nu_{1}^{(n)}(t_{1})\nu_{1}^{(n)}(t_{2})^{*} + ~\nu_{2}^{(n)}(t_{1})\nu_{2}^{(n)}(t_{2})^{*}\right]
      + \hat{b}^{\dagger}{}^{2}~\nu_{2}^{(n)}(t_{1})\nu_{1}^{(n)}(t_{2})^{*} + \hat{b}^{2}~\nu_{1}^{(n)}(t_{1})\nu_{1}^{(n)}(t_{2})^{*}\\
    & + \hat{b}^{\dagger}\left[\nu_{1}^{(n)}(t_{1})\nu_{3}^{(n)}(t_{2})^{*} + \nu_{3}^{(n)}(t_{1})\nu_{2}^{(n)}(t_{2})^{*}\right]
      + \hat{b}\left[\nu_{2}^{(n)}(t_{1})\nu_{3}^{(n)}(t_{2})^{*} + \nu_{3}^{(n)}(t_{1})\nu_{1}^{(n)}(t_{2})^{*}\right]\\
    & + \nu_{2}^{(n)}(t_{1})\nu_{2}^{(n)}(t_{2})^{*} + \nu_{3}^{(n)}(t_{1})\nu_{3}^{(n)}(t_{2})^{*}\Big]\ket{\psi(0)}.
\end{aligned}
\end{equation}

Looking at the structure of the products of $\nu_{l}^{(n)}(t_{m})$, we can assert that the fundamental need is to calculate the integrals
\begin{equation}
    \int_{0}^{t}~dt_{1}~e^{(\Gamma - \mi\omega)t_{1}}~\int_{0}^{t}~dt_{2}~e^{(\Gamma + \mi\omega)t_{2}}~\nu_{j}^{(n)}(t_{1})\nu_{k}^{(n)}(t_{2}),\quad j,k\in\left\{1, 2, 3\right\}.
\end{equation}
But, if we look at the index condition $j = k\in\left\{1, 2, 3\right\}\equiv l$, the relation is reduced to
\begin{equation}
    \Bigg\vert\int_{0}^{t}~d\tau~e^{(\Gamma - \mi\omega)\tau}~\nu_{l}^{(n)}(\tau)\Bigg\vert^{2},
\end{equation}
denoting that in final instance, the evaluation of three integrals is all needed at this point, explicitly given as an operator functional of $\hat{n}$ by
\begin{equation}\label{eq:EW_coeffs}
\begin{aligned}
    \int_{0}^{t}~d\tau~e^{(\Gamma - \mi\omega)\tau}~\nu_{1}^{(\hat{n})}(\tau) &= \frac{e^{t\left(\gamma  E_{\gamma}(\hat{n}) + (\Gamma - \mi\omega)\right)} - 1}{\gamma  E_{\gamma}(\hat{n}) + (\Gamma - \mi\omega)}\cosh^{2}r(\hat{n}) - \frac{e^{t\left(\gamma  E_{\gamma}(\hat{n})^{*} + (\Gamma - \mi\omega)\right)} - 1}{\gamma  E_{\gamma}(\hat{n})^{*} + (\Gamma - \mi\omega)}\sinh^{2}r(\hat{n}) \equiv \mathcal{L}_{1}^{(\hat{n})}(t)\\
    \int_{0}^{t}~d\tau~e^{(\Gamma - \mi\omega)\tau}~\nu_{2}^{(\hat{n})}(\tau) &= \frac{1}{2}\left[\frac{e^{t\left(\gamma  E_{\gamma}(\hat{n}) + (\Gamma - \mi\omega)\right)} - 1}{\gamma  E_{\gamma}(\hat{n}) + (\Gamma - \mi\omega)} - \frac{e^{t\left(\gamma  E_{\gamma}(\hat{n})^{*} + (\Gamma - \mi\omega)\right)} - 1}{\gamma  E_{\gamma}(\hat{n})^{*} + (\Gamma - \mi\omega)}\right]\sinh 2r(\hat{n}) \equiv \mathcal{L}_{2}^{(\hat{n})}(t)\\
    \int_{0}^{t}~d\tau~e^{(\Gamma - \mi\omega)\tau}~\nu_{3}^{(\hat{n})}(\tau) &= \alpha(\hat{n})\left[\frac{e^{t\left(\gamma  E_{\gamma}(\hat{n}) + (\Gamma - \mi\omega)\right)} - 1}{\gamma  E_{\gamma}(\hat{n}) + (\Gamma - \mi\omega)}\cosh r(\hat{n}) - \frac{e^{t\left(\gamma  E_{\gamma}(\hat{n})^{*} + (\Gamma - \mi\omega)\right)} - 1}{\gamma  E_{\gamma}(\hat{n})^{*} + (\Gamma - \mi\omega)}\sinh r(\hat{n})\right.\\
    &\left.\hspace{3.15em} - \frac{e^{t (\Gamma - \mi\omega)} - 1}{\Gamma - \mi\omega}e^{-r(\hat{n})}\right] \equiv \mathcal{L}_{3}^{(\hat{n})}(t)
\end{aligned}
\end{equation}
writing $E_{\gamma}(\hat{n}) = e^{\frac{\mi\tilde{\omega}_{m}(\hat{n})}{\gamma}} - 1$, which carried the damping effects of the decoherence. The cross-terms, $j\neq k$, can be evaluated using these previous results. Looking at the denominators, there's a play between the $E_{\gamma}$ term, the proper decoherence rate $\gamma$ and the displacement of the spectrum. Thus, the position of the spectral peaks of the sidebands will be displaced proportional to how much decoherence is at the beginning. On the other hand, a single term in $\mathcal{L}_{3}^{(n)}(t)$, that doesn't depend on these previous quantities, will survive long-time evolutions, leading to a single spectral peak, relatable to the cavity field contribution, as a function of the couplings $g_{l}$ and $g_{q}$. Finally, note that the spectrum doesn't depend on the frequency $\omega_{c}$ of the cavity field.

At this point, the initial condition will be taken as a number state in the field, and a coherent state in the mirror $\ket{\psi(0)} = \ket{n}_{c}\otimes\ket{\beta}_{m}$, and then generalized to $\ket{\psi(0)} = e^{-\tilde{\alpha}^{2}}\sum_{k = 0}^{\infty}(\tilde{\alpha}^{2k}/k!)~\ket{n}_{c}\otimes\ket{\beta}_{m} \equiv \ket{\tilde{\alpha}, \beta}$, this because the interest to evaluate single photon excitations reflecting on the mirror without initial phonons, and the superposition of photonic and phononic distributions in both. Note that the relations \eqref{eq:EW_coeffs} are suitable to be used with arbitrary initial condition, since it depends on the photonic operator $\hat{n}$. Then, we are now in position to obtain the time-dependent spectral response of the mirror, which is 
\begin{equation}\label{eq:S}
\begin{aligned}
    \mathcal{S}^{(n,\beta)}(\Gamma, \omega; t) = 2\Gamma e^{-2\Gamma t}&\Big[\vert\beta\vert^{2}\left(\vert\mathcal{L}_{1}^{(n)}(t)\vert^{2} + \vert\mathcal{L}_{2}^{(n)}(t)\vert^{2}\right)
     + \beta^{*}{}^{2}~\mathcal{L}_{2}^{(n)}(t)\mathcal{L}_{1}^{(n)}(t)^{*} + \beta^{2}~\mathcal{L}_{1}^{(n)}(t)\mathcal{L}_{2}^{(n)}(t)^{*}\\
     & + \beta^{*}\left[\mathcal{L}_{1}^{(n)}(t)\mathcal{L}_{3}^{(n)}(t)^{*} + \mathcal{L}_{3}^{(n)}(t)\mathcal{L}_{2^{(n)}}(t)^{*}\right]
     + \beta\left[\mathcal{L}_{2}^{(n)}(t)\mathcal{L}_{3}^{(n)}(t)^{*} + \mathcal{L}_{3}^{(n)}(t)\mathcal{L}_{1}^{(n)}(t)^{*}\right]\\
    & + \vert\mathcal{L}_{2}^{(n)}(t)\vert^{2} + \vert\mathcal{L}_{3}^{(n)}(t)\vert^{2}\Big],
\end{aligned}
\end{equation}
for mean number of photons $\bar{n}$, and mean coherent phonons in the mirror $\vert\beta\vert^{2} = \bar{N}$. When a coherent photon distribution $\ket{\tilde{\alpha}}$ is requested, it can be calculated as
\begin{equation}
    \mathcal{S}^{(\tilde{\alpha}, \beta)} = e^{-\tilde{\alpha}^{2}}\sum\limits_{k = 0}^{\infty}\frac{\tilde{\alpha}^{2k}}{k!}\,\mathcal{S}^{(n,\beta)}(\Gamma, \omega; t),
\end{equation}
were abusing the notation, we use $\tilde{\alpha}$ as the coherent state in the cavity field, which not to be confused with the $\alpha(r(\hat{n}))$ form of the displacement operator in the diagonalization.

\begin{figure}[htbp]
    \centering
    \includegraphics[width = \linewidth]{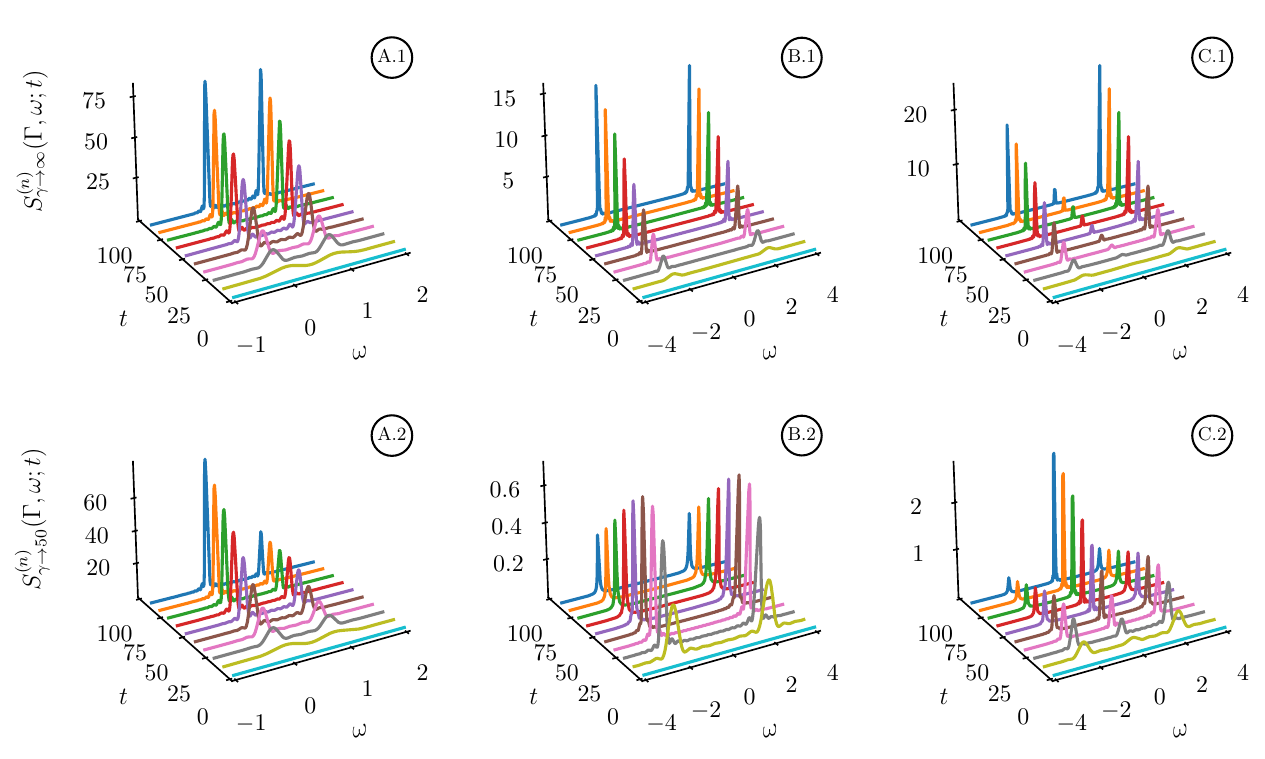}
    \caption{Time-dependent spectrum $\mathcal{S}_{\gamma}(\Gamma, \omega; t)$ for the moving mirror. The upper row $\square.1$ are evolutions without decoherence, $\gamma\rightarrow\infty$, whereas each column, $A.1$, $B.2$ and $C.3$, represents linear, quadratic and linear-quadratic coupling, respectively. The same labeling system for the lower row. Here the relevant parameters are, $\Gamma = 0.01$, $\omega_{c}=\omega_{m}=1.0$, $g_{l}=g_{q}=1.0$. The initial state is one photonic excitation in the field, no phonons in the moving mirror, $\ket{\psi(0)}_{\mathrm{num}} = \ket{1,0}$.}
    \label{fig:S_num}
\end{figure}

\begin{figure}[htbp]
    \centering
    \includegraphics[width = \linewidth]{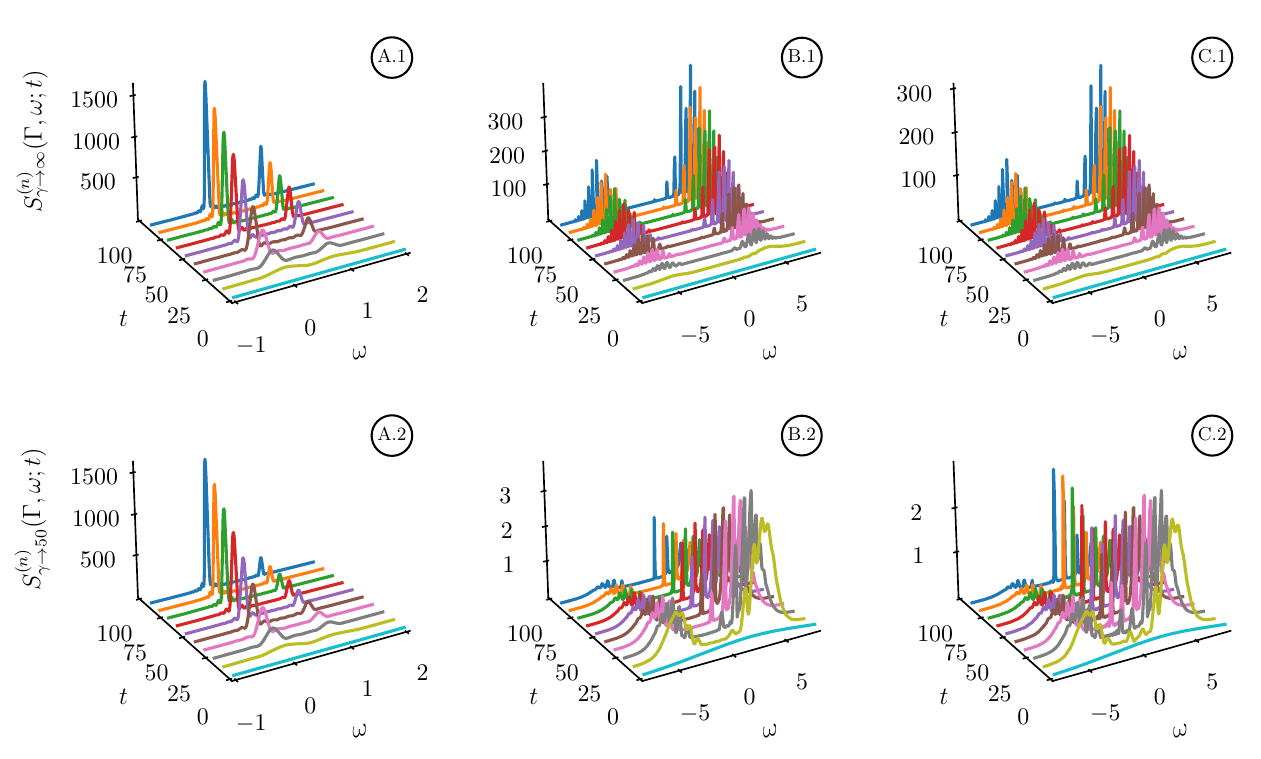}
    \caption{Time-dependent spectrum $\mathcal{S}_{\gamma}(\Gamma, \omega; t)$ for the moving mirror. The upper row $\square.1$ are evolutions without decoherence, $\gamma\rightarrow\infty$, whereas each column, $A.1$, $B.2$ and $C.3$, represents linear, quadratic and linear-quadratic coupling, respectively. The same labeling system for the lower row. Here the relevant parameters are, $\Gamma = 0.01$, $\omega_{c}=\omega_{m}=1.0$, $g_{l}=g_{q}=1.0$. The initial state is a coherent state in the cavity field and in the moving mirror, $\ket{\psi(0)}_{\mathrm{coh}} = \ket{2,2}$.}
    \label{fig:S_coh}
\end{figure}

In Figures~\ref{fig:S_num} and \ref{fig:S_coh} it is shown the evolution of the spectrum as a function of time, comparing when there's no decoherence $(\gamma\rightarrow\infty)$, upper row with labels $\square.1$, and when there's decoherence enabled $(\gamma \ll \infty)$, lower row with $\square.2$; columns labels denoted if there's linear $A.\square$, quadratic $B.\square$, or linear and quadratic $C.\square$ coupling acting. Starting in Fig.~\ref{fig:S_num}, we have an initial state $\ket{\psi(0)}_{\mathrm{num}} = \ket{1,0}$, intended to show the response of the mirror with a single excitation in the field. The first column, where there's only linear coupling, has spectral peaks centered at $\omega = \{0, \gamma\sin\tfrac{\omega_{m}}{\gamma}\}$; when there's no decoherence $(A.1)$, the peaks have equal amplitude through the evolution, but when the decoherence is active $(A.2)$, the peak at $\omega = \gamma\sin\tfrac{\omega_{m}}{\gamma}$ is diminished in time, prevailing just the peak at $\omega = 0$, indicating that the mirror stops converting radiation pressure in phonons. Because $g_{q} = 0$, the only terms playing are $\mathcal{L}_{1}^{(n)}(t)$ and $\mathcal{L}_{3}^{(n)}(t)$. Moving to the second columns, where there's only linear coupling, there are peaks centered at $\omega = \pm \gamma\sin\frac{\bar{\omega}_{m}(n)}{\gamma}$, note there's a contribution of the modified frequency $\bar{\omega}_{m}(n)$ which displaces the peak proportional to the number of photons present in the cavity. Here, because $g_{l} = 0$, the terms that contribute are $\mathcal{L}_{1}^{(n)}(t)$ and $\mathcal{L}_{2}^{(n)}(t)$. The decoherence evolution shows no central peak at $\omega = 0$, it means that both peaks' amplitude will continue damping until it disappear. Finally, the third column has enabled the linear and quadratic coupling simultaneously, here the spectral peaks are the combination of the previous discussed, $\omega = \{0, \pm\gamma\sin\frac{\bar{\omega}_{m}(n)}{\gamma}\}$, meaning that the central peak at $\omega = 0$ will be the most prominent in time when the decoherence takes places, which can be interpreted as the mirror is cooling down, evolving with a constant mean phonon value $\bar{N}$, so no more energetic transitions will be displayed in the spectrum. In Fig.~\ref{fig:S_coh} it is shown the spectrum in the same parameters conditions, but starting with coherent states, in the field and the mirror, $\ket{\psi(0)}_{\mathrm{coh}} = \ket{2, 2}$. The spectral responses are similar as the previous discussed, but observing multipeaks transitions when there's quadratic and linear-quadratic couplings. But as with the single photonic excitation, the fate of the evolution when there's decoherence, is to develop a central peak at $\omega = 0$, denoting that the conversion of photon-phonon is stopping.

\paragraph*{Long-time limit.} It is worth calculating the fate of the evolution as decoherence takes place. For this, we need to know the limits $\lim\limits_{t\rightarrow\infty} e^{-2\Gamma t}\vert\mathcal{L}_{k}^{(n)}(t)\vert^{2}$, and $\lim\limits_{t\rightarrow\infty} e^{-2\Gamma t}\mathcal{L}_{j}^{(n)}(t)\mathcal{L}_{k}^{(n)}(t)^{*}$. Notice that these coefficients defined in \eqref{eq:EW_coeffs}, the only terms that doesn't depend on $E_{\gamma}(n)$ will be preserved, which is existing in the last term of $\mathcal{L}_{3}^{(n)}(t)$. This will lead us to the long-time spectrum of the form,
\begin{equation}\label{eq:S_longtime}
    \lim\limits_{t \rightarrow \infty} \mathcal{S}^{(n,\beta)}(\Gamma, \omega; t) = \frac{2\Gamma}{\Gamma^{2} + \omega^{2}}\alpha(n)e^{-r(n)}.
\end{equation}
This expression tells us that there's only one spectral peak at $\omega = 0$, and that this peak only appears when there's either linear or linear-quadratic  because when there's only quadratic coupling, $\alpha(n) = 0$, meaning we obtain no spectral signature at final, as we can preview in panels $\mathrm{(B.2)}$ of Figs.~\ref{fig:NXnum_expect}-\ref{fig:NXcoh_expect}. Moreover, as we can see, the final spectral form doesn't depend on the initial state of the moving mirror, only on what we have in the cavity field at the beginning, $\ket{n}$ or $\ket{\tilde{\alpha}}$.

\section{Conclusions}\label{sec:conclusions}
In this manuscript, the description of the dynamics and the spectral response in the linear-quadratic moving mirror-field system under pure dephasing decoherence, is addressed. From the operational side, the full diagonalization using unitary transformations was done, since all the algebra generators closed under the Lie parenthesis, full analytical solutions can be obtained. The diagonalized Hamiltonian is composed of displaced and squeezed contributions, which leads to non-linear phenomena like revival-like of mean values when quadratic couplings are considered. The pure dephasing decoherence is introduced in the framework of the Milburn's equation, which can be reduced to the Schr\"odinger-Heisenberg evolution when the decoherence rate is carried as an infinite limit. Finally, the time-dependent spectrum, for the mirror response, was obtained using the Eberly-W\'odkiewicz non-stationary spectrum. Emphasize is given to the response of the mirror when there's a single photonic excitation, and no phonons. It is shown how the well-known results for the linear only coupling is modified by the inclusion of the second order term, the quadratic coupling, plus a view in the only quadratic is also given for sake of comparison complement. From the dynamical standpoint, the evolution, considering decoherence enabled, shows diminishing in the amplitudes of the mean values of phonons and the position quadrature. It is worth noticing that in a quadratic only coupling, it's equivalent to a mirror not moving at all. We can understand this as because the response of the mirror is to convert the photon to phonons due to radiation pressure, but the dephasing proportional to the decoherence rate $\gamma$, damped this behavior, leading to a mean value in the phonons $\bar{N}$ and the quadrature $\bar{X}$, for a critical time $t_{c}$ where's there's no more photon-phonon conversion. From the time-dependent spectral standpoint, the response of the mirror shows, for linear only coupling, that there are energetic transitions between the sidebands $\omega = 0, \omega_{m}$, leading to just a single spectral peak at $\omega = 0$, when the decoherence takes places; on the other hand, when there's quadratic only coupling, the unique chance the mirror has is to stop their movement, the transitions occur at $\omega = \pm\gamma\sin\tfrac{\bar{\omega}_{m}(n)}{\gamma}$, but disappear at long-time; finally, when there are both couplings, linear and quadratic, it is expected to see transitions occurring as a triplet at $\omega = 0, \pm\gamma\sin\tfrac{\bar{\omega}_{m}(n)}{\gamma}$, which carries the signatures of both contributions. Long-time limits were also presented, showing that only a spectral peak at $\omega_{0}$ is expected in linear and linear-quadratic couplings, and no spectral peak is expected in quadratic only coupling. The same analysis can be done when coherent states in both, the field and the mirror, are prepared at the beginning. In this case, multipeaks are present, indicating simultaneous transitions. At the long-time limit, it is observed the same as the single photonic excitation, a spectral peak at $\omega = 0$, or nothing at all, indicating the mirror stops its photon-phonon conversion.

\section*{Acknowledgements}
A.R.-U. acknowledges financial support by UNAM Posdoctoral Program (POSDOC) 2024-2025, and to ICF-UNAM for the assistance in-place. The author thanks F. R\'ecamier-Angelini for his guidance through fruitful discussions about optomechanical systems. We also thank Reyes Garc\'ia (C\'omputo-ICF) for maintaining our computing servers.

\appendix
\section{Diagonalization of the Hamiltonian}\label{app:diag}
We now present the diagonalization of the Hamiltonian using unitary transformations. The Hamiltonian, as a reminder, is defined in full form by
\begin{equation}\label{H}
	\hat{H} = \omega_{c}\hat{n} + \omega_{m}\hat{N} - g_{l}\hat{n}\left(\hat{b}^{\dagger} + \hat{b}\right) + g_{q}\hat{n}\left(\hat{b}^{\dagger} + \hat{b}\right)^{2},
\end{equation}
representing and optomechanical system whose mechanical frequency depending on the position is expanded up-to second order. Inspecting the interaction terms, with linear $g_{l}$ and quadratic $g_{q}$ coupling, we advert that geometrical unitary transformations must be needed: first, a one-mode squeezing, and second, a one-mode displacement in the mechanical operators.
\paragraph*{One-mode squeezing.} The one-mode squeezing transformation is defined by the operator
\begin{equation}
	\hat{S}(r(\hat{n})) = e^{\frac{1}{2}r(\hat{n})\left(\hat{b}^{\dagger}{}^{2} - \hat{b}^{2}\right)},\qquad \hat{S}^{\dagger}(r(\hat{n}))\equiv \hat{S}(-r(\hat{n})),
\end{equation}
where $r(\hat{n})$ is a function depending on the number of photons, to be determined. The action of this transformation on the mirror operational basis is
\begin{equation}
\begin{aligned}
    \hat{S}(r(\hat{n}))\,\hat{b}^{\dagger}\,\hat{S}^{\dagger}(r(\hat{n})) &= \hat{b}^{\dagger}\cosh r(\hat{n}) - \hat{b}\sinh r(\hat{n}),\\
    \hat{S}(r(\hat{n}))\,\hat{b}\,\hat{S}^{\dagger}(r(\hat{n})) &= \hat{b}\cosh r(\hat{n}) - \hat{b}^{\dagger}\sinh r(\hat{n}).
\end{aligned}
\end{equation}
Knowings this, we can take the full transformation of the Hamiltonian \eqref{H}, resulting in
\begin{equation}
\begin{aligned}
    \hat{S}(r)\hat{H}\hat{S}^{\dagger}(r) &= \omega_{c}\hat{n} + \omega_{m}\hat{N}\cosh 2r - g_{l}\hat{n}\left(\hat{b}^{\dagger} + \hat{b}\right)e^{-r}\\ 
    &- \frac{\omega_{m}}{2}\left(\hat{b}^{\dagger}{}^{2} + \hat{b}^{2}\right)\sinh 2r + g_{q}\hat{n}\left(\hat{b}^{\dagger} + \hat{b}\right)^{2}e^{-2r},
\end{aligned}
\end{equation}
which can be rearranged to highlight the quadratic terms we want to eliminate. The equation and the elimination conditions are, respectively, 
\begin{equation}
	\begin{aligned}
		&\left(-\frac{\omega_{m}}{2}\sinh 2r + g_{q}\hat{n}e^{-2r}\right)\left(\hat{b}^{\dagger}{}^{2} + \hat{b}^{2}\right) = 0\\
		&\Rightarrow r(\hat{n}) = \frac{1}{4}\ln\left\vert 1 + \frac{4g_{q}\hat{n}}{\omega_{m}}\right\vert,
	\end{aligned}
\end{equation}
where $r(\hat{n})$ is now identified as the squeezing parameter, preserving the unitariness of the transformation. The Hamiltonian after this one-mode squeezing operation takes the form
\begin{equation}\label{Hs}
    \hat{H}_{\texttt{S}} = \omega_{c}\hat{n} + \bar{\omega}_{m}(\hat{n})\hat{N} - g_{l}\hat{n}\left(\hat{b}^{\dagger} + \hat{b}\right)e^{-r},\qquad  \bar{\omega}_{m}(\hat{n}) = \sqrt{\omega_{m}\left(\omega_{m} + 4g_{q}\hat{n}\right)},
\end{equation}
where the new frequency $\bar{\omega}_{m}(\hat{n})$ can be identified as a photon number-dependent frequency in the mechanical part influenced by the photonic counterpart.

\paragraph*{One-mode displacement.} Because the one-mode squeezing transformation only deals with the quadratic part of the Hamiltonian \eqref{H}, the resulting Hamiltonian \eqref{Hs} still has the lineal part damped with an exponential term in function of $r(\hat{n})$. A displacement transformation can be used, defined by
\begin{equation}
	\hat{D}(\alpha(\hat{n})) = e^{\alpha(\hat{n})\left(\hat{b}^{\dagger} - \hat{b}\right)},\qquad \hat{D}^{\dagger}(\alpha)\equiv\hat{D}(-\alpha),
\end{equation}
whose action on the mechanical operators is
\begin{equation}\label{eq:D's}
	\begin{aligned}
		\hat{D}(\alpha(\hat{n}))\hat{b}^{\dagger}\hat{D}^{\dagger}(\alpha(\hat{n})) &= \hat{b}^{\dagger} - \alpha(\hat{n}),\\
		\hat{D}(\alpha(\hat{n}))\hat{b}\hat{D}^{\dagger}(\alpha(\hat{n})) &= \hat{b} - \alpha(\hat{n}).
	\end{aligned},
\end{equation}
where the displacement argument observes realness, $\im\left\{\alpha(\hat{n})\right\} = 0$.

Transforming the \emph{squeezed} Hamiltonian \eqref{Hs} using the rules \ref{eq:D's}, we can identify the relations 
\begin{equation}
	\begin{aligned}
		\hat{D}(\alpha)\left(\omega_{c}\hat{n} + \bar{\omega}_{m}(\hat{n})\hat{N}\right)\hat{D}^{\dagger}(\alpha)& = \omega_{c}\hat{n} + \bar{\omega}_{m}(\hat{n})\hat{N} - \bar{\omega}_{m}(\hat{n})\alpha\left(\hat{b}^{\dagger} + \hat{b}\right) + \bar{\omega}_{m}(\hat{n})\vert\alpha\vert^{2} = \hat{H}_{\texttt{SD}}\\
		&\equiv \omega_{c}\hat{n} + \bar{\omega}_{m}(\hat{n})\hat{N} - g_{l}\hat{n}\left(\hat{b}^{\dagger} + \hat{b}\right)e^{-r}\\
		&\Rightarrow \alpha(\hat{n}) = -\frac{g_{l}\hat{n}e^{-r}}{\bar{\omega}_{m}(\hat{n})},\qquad \therefore~\bar{\omega}_{m}(\hat{n})\vert\alpha(\hat{n})\vert^{2} = \frac{g_{l}^{2}\hat{n}^{2}e^{-2r}}{\bar{\omega}_{m}(\hat{n})},
	\end{aligned}	
\end{equation}
which define completely the transformation and the coefficient $\alpha(\hat{n})$. The full diagonal Hamiltonian now takes the form
\begin{equation}\label{Hsd}
	\hat{H}_{\texttt{diag}} = \omega_{c}\hat{n} + \bar{\omega}_{m}(\hat{n})\hat{N} - \frac{g_{l}^{2}\hat{n}^{2}e^{-2r}}{\bar{\omega}_{m}(\hat{n})},
\end{equation}
where it can be identified as two harmonic oscillators, influenced by a Kerr-like term in the photonic field due to the presence of $\hat{n}^{2}$.

\paragraph*{Original representation.} The diagonal representation in the original reference frame of the Hamiltonian \eqref{H} is the successive transformation by $\hat{S}(r(\hat{n}))$ and $\hat{D}(\alpha(\hat{n}))$,
\begin{equation}\label{eq:Usd}
	\begin{aligned}
		\hat{H}_{\texttt{diag}} &\equiv \hat{D}(\alpha(\hat{n}))\hat{S}(r(\hat{n}))\hat{H}\hat{S}^{\dagger}(r(\hat{n}))\hat{D}^{\dagger}(\alpha(\hat{n}))\\
		&\Rightarrow \hat{H} = \hat{S}^{\dagger}(r(\hat{n}))\hat{D}(\alpha(\hat{n}))\,\hat{H}_{\texttt{diag}}\,\hat{D}(\alpha(\hat{n}))\hat{S}(r(\hat{n})).
	\end{aligned}
\end{equation}

\paragraph*{Heisenberg representation and expectation values.} Relation \eqref{eq:Usd} defines the evolution operator of the full system
\begin{equation}\label{U}
	\hat{\mathcal{U}}(t) = \hat{S}^{\dagger}(r)\hat{D}(\alpha)e^{-\mi t\hat{H}_{\texttt{diag}}}\hat{D}(\alpha)\hat{S}(r),
\end{equation}
that is used to calculate the time evolution of operators. The Heisenberg representation of the mechanical operators is obtained as
\begin{equation}
	\begin{aligned}
		\hat{U}^{\dagger}(t)\hat{b}^{\dagger}\hat{U}(t) &= \xi_{1}(t)\hat{b}^{\dagger} + \xi_{2}(t)\hat{b} + \xi_{3}(t),\\
		\hat{U}^{\dagger}(t)\hat{b}\,\hat{U}(t) &= \xi_{1}^{*}(t)\hat{b} + \xi_{2}^{*}(t)\hat{b}^{\dagger} + \xi_{3}^{*}(t)
	\end{aligned}
\end{equation}
where the companion time-dependent coefficients are
\begin{equation}\label{eq:exp_coeffs}
	\begin{aligned}
		\xi_{1}(t) &= \cos\left(\bar{\omega}_{m}(\hat{n}) t\right) + \mi\cosh\left(2r\right)\sin\left(\bar{\omega}_{m}(\hat{n}) t\right),\\
		\xi_{2}(t) &= \mi\,\sinh\left(2r\right)\sin\left(\bar{\omega}_{m}(\hat{n}) t\right),\\
		\xi_{3}(t) &= \alpha\left[\left(1 - e^{\mi\bar{\omega}_{m}(\hat{n}) t}\right)\cosh\left(r\right) + (1 - e^{-\mi\bar{\omega}_{m}(\hat{n}) t})\sinh\left(r\right)\right],\\
	\end{aligned}
\end{equation}
determining completely the dynamics of the time evolution.

\bibliography{refs}
\bibliographystyle{apsrev4-2}

\end{document}